\documentclass[10pt,twocolumn,letterpaper]{article}
\pdfoutput=1

\usepackage{cvpr}
\usepackage{times}
\usepackage{epsfig}
\usepackage{graphicx}
\usepackage{amsmath}
\usepackage{amssymb}
\usepackage{float}
\usepackage{subcaption}
\usepackage{wrapfig}
\usepackage[export]{adjustbox}

\usepackage[breaklinks=true,bookmarks=false]{hyperref}

\cvprfinalcopy 

\def\cvprPaperID{20} 

\begin{document}

\title{MediaPipe: A Framework for Building Perception Pipelines}

\author{
Camillo Lugaresi, Jiuqiang Tang, Hadon Nash, Chris McClanahan, Esha Uboweja, Michael Hays,\\
Fan Zhang, Chuo-Ling Chang, Ming Guang Yong, Juhyun Lee, Wan-Teh Chang, Wei Hua,\\
Manfred Georg and Matthias Grundmann\\
Google Research\\
{\tt\small mediapipe@google.com}\\
}

\maketitle

\begin{abstract}

Building applications that perceive the world around them is challenging.
A developer needs to (a) select and develop corresponding machine learning algorithms and models, (b) build a series of prototypes and demos, (c) balance resource consumption against the quality of the solutions, and finally (d) identify and mitigate problematic cases.
The MediaPipe framework addresses all of these challenges.
A developer can use MediaPipe to build prototypes by combining existing perception components, to advance them to polished cross-platform applications and measure system performance and resource consumption on target platforms.
We show that these features enable a developer to focus on the algorithm or model development and use MediaPipe as an environment for iteratively improving their application with results reproducible across different devices and platforms.
MediaPipe will be open-sourced at \small{\url{https://github.com/google/mediapipe}}.
\end{abstract}


\section{Introduction}

\begin{figure}[t!] 
\begin{center}
   \includegraphics[width=0.85\linewidth]{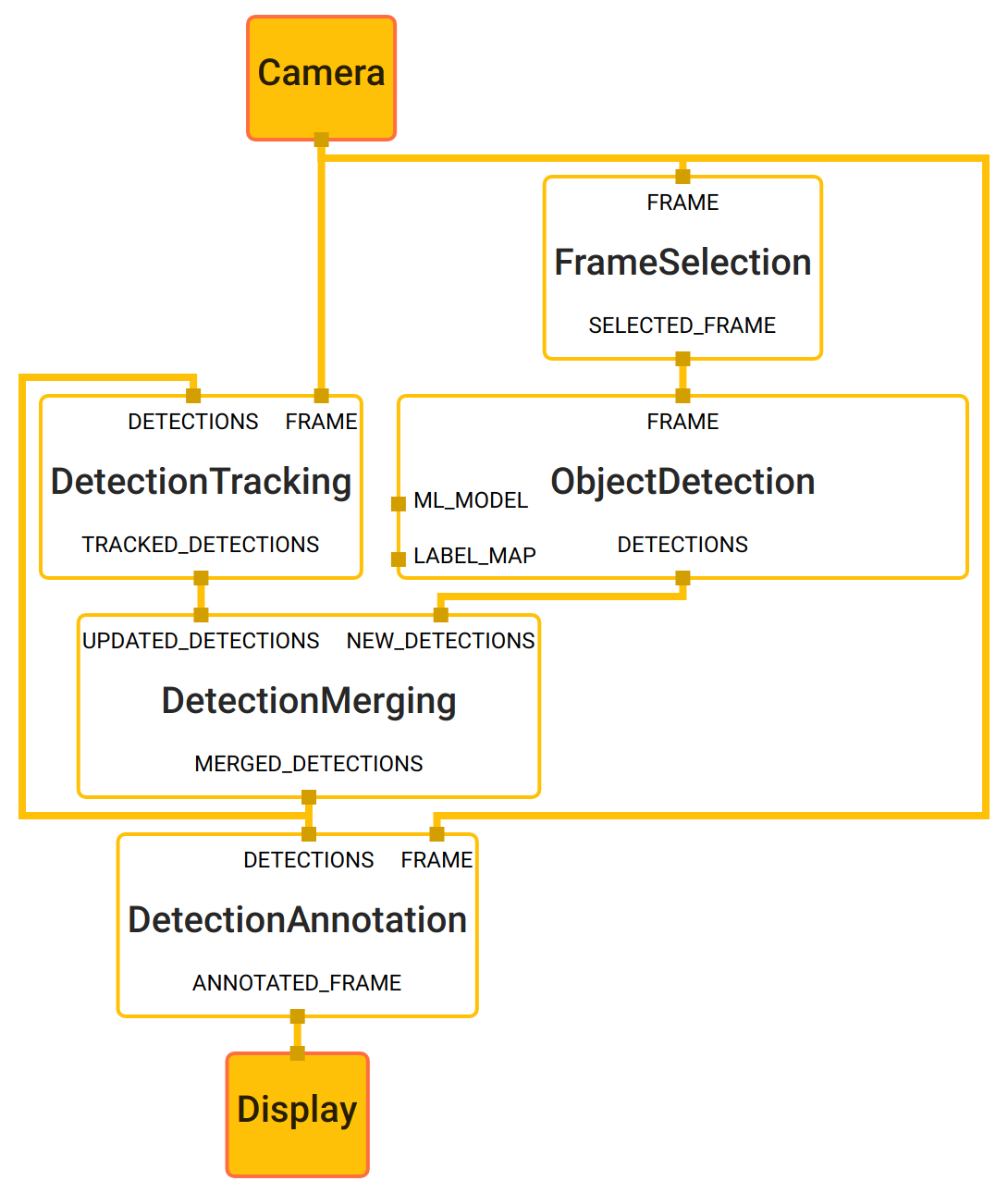}
\end{center}
   \caption{Object detection using MediaPipe. The transparent boxes represent computation nodes (calculators) in a MediaPipe graph, solid boxes represent external input/output to the graph, and the lines entering the top and exiting the bottom of the nodes represent the input and output streams respectively. The ports on the left of some nodes denote the input side packets. See Section \ref{object} for details.}
\label{fig:objectgraph}
\end{figure}

MediaPipe is a framework for building pipelines to perform inference over arbitrary sensory data.
With MediaPipe, a \textit{perception pipeline} can be built as a graph of modular components, including model inference, media processing algorithms and data transformations, etc. Sensory data such as audio and video streams enter the graph, and perceived descriptions such as object-localization and face-landmark streams exit the graph. An example is shown in Figure~\ref{fig:objectgraph}.

MediaPipe is designed for machine learning (ML) practitioners, including researchers, students, and software developers, who implement production-ready ML applications, publish code accompanying research work, and build technology prototypes. The main use case for MediaPipe is rapid prototyping of perception pipelines with inference models and other reusable components. MediaPipe also facilitates the deployment of perception technology into demos and applications on a wide variety of different hardware platforms. MediaPipe enables incremental improvements to perception pipelines through its rich configuration language and evaluation tools.

Modifying a perception application to incorporate additional processing steps or inference models can be difficult, due to excessive coupling between steps.
In addition, developing the same application for different platforms is time consuming and usually involves optimizing inference and processing steps to run correctly and efficiently on a target device.

MediaPipe addresses these challenges by abstracting and connecting individual perception models into maintainable pipelines. All of the steps necessary to infer from the sensory data and get the perceived results are specified in the pipeline configuration. It is easy to re-use MediaPipe components in different pipelines across successive applications as these components share a common interface oriented around time-series data. 
Each pipeline can then run with the same behavior on a variety of platforms, enabling the practitioner to develop the application on workstations, and then deploy it on mobile, for example.

MediaPipe consists of three main parts: (a) a framework for inference from sensory data, (b) a set of tools for performance evaluation, and (c) a collection of re-usable inference and processing components called calculators. We describe the framework in Sections \ref{architecture} and \ref{implementation}, and the set of tools in Section \ref{tools}. We also present example MediaPipe perception applications in Section \ref{application}. 



\section{Related work}

Media analysis is an active area of research in both academia and industry. Typically, a media file or camera input containing both audio and video is extracted into separate streams via a media decoder which are then analyzed separately. Neural net engines like TensorFlow~\cite{tf16}, PyTorch~\cite{pytorch}, CNTK~\cite{cntk}, MXNet~\cite{mxnet} represent their neural networks in forms of directed graphs whose nodes are simple and deterministic, \ie one input generates one output, which allows very efficient execution of the compute graph consisting of such lower level semantics. MediaPipe on the other hand, operates at much higher level semantics and allows more complex and dynamic behavior, \ie one input can generate zero, one or multiple outputs, which cannot be modeled with neural networks. This complexity allows MediaPipe to excel at analyzing media at higher semantics.

Other forms of graph processing and data flow frameworks such as Beam~\cite{beam} and Dataflow~\cite{dataflow} operate on clusters of compute machines.  While the dependency of each operation is defined in the form of graphs as well, Beam~\cite{beam} and Dataflow~\cite{dataflow} handle big chunks of data in a batching fashion rather than in a streaming fashion, which makes them unsuitable for the audio/video processing domain.

The research project Ptolemy~\cite{ptolemy} studies concurrent systems, but it heavily focuses on modeling and simulation for the purpose of studying such systems.
With MediaPipe, a developer can build and analyze concurrent systems via graphs, and further deploy such systems as performant applications. However, MediaPipe is targeted towards applications in the audio/video processing domain and not limited to the scope of modeling the performance of concurrent systems.

In the field of robotics, ROS~\cite{ros} allows the definition of the processing logic in the form of graphs, where each node is a process, communicating with other nodes via inter-process communication (IPC) calls. With MediaPipe, we avoid the additional overhead and complexity of IPC calls. 

GStreamer~\cite{gs01} is  a framework for  constructing  arbitrary  graphs  of  media-handling  components with low-level operations.  However, GStreamer primarily targets audio/video  media  editing rather than analysis.
MediaPipe with its higher level semantics makes it more suitable audio/video analysis and understanding.

OpenCV 4.0 introduced the Graph API (G-API)~\cite{cv18} which allows specification of sequences of OpenCV image processing operations in the form of a graph. In contrast, our framework allows operations on arbitrary data types and has native support for streaming time-series data which makes it much more suitable for analyzing audio and sensor data.

\section{Architecture} \label{architecture}

MediaPipe allows a developer to prototype a pipeline incrementally. A pipeline is defined as a directed graph of components where each component is a \texttt{Calculator}. The graph is specified using a \texttt{GraphConfig} protocol buffer and then run using a \texttt{Graph} object. 

In the graph, the calculators are connected by data \texttt{Streams}. Each stream represents a time-series of data \texttt{Packets}. Together, the calculators and streams define a data-flow graph. The packets which flow across the graph are collated by their timestamps within the time-series.

The pipeline can be refined incrementally by inserting or replacing calculators anywhere in the graph. Developers can also define custom calculators. While the graph executes calculators in parallel, each calculator executes on at most one thread at a time\footnote{An advanced feature enables parallel execution of a calculator on several threads assuming temporal independence of the time-series}. This constraint, coupled with the immutability of data packets, ensures that custom calculators can be defined without specialized expertise in multi-threaded programming.

\subsection{Packets}

The basic data unit in MediaPipe is a \texttt{Packet}. A packet consists of a numeric timestamp and a shared pointer to an immutable payload. The payload can be of any C++ type, and the payload's type is also referred to as the type of the packet. Packets are treated as value classes and can be copied cheaply. Each copy shares ownership of the payload, with reference-counting semantics. Each copy has its own timestamp.

\subsection{Streams}
Each node in the graph is connected to another node through a \texttt{Stream}. A stream carries a sequence of packets whose timestamps must be monotonically increasing. An output stream can be connected to any number of input streams of the same type. Each input stream receives a separate copy of the packets from an output stream, and maintains its own queue to allow the receiving node to consume the packets at its own pace.

\subsection{Side packets}
A side-packet connection between nodes carries a single packet with an unspecified timestamp. It can be used to provide some data that will remain constant, whereas a stream represents a flow of data that changes over time. For example, the string defining the file path for an ML model can be fed into a node through a side packet.

\subsection{Calculators} \label{calculators}
Each node in the graph is implemented as a \texttt{Calculator}. The bulk of graph execution happens inside its calculators. A calculator may receive zero or more input streams and/or side packets and produces zero or more output streams and/or side packets.

Each calculator included in a program is registered with the framework so that the graph configuration (further discussed in Section \ref{graph_config}) can reference it by name.

Calculators are highly customizable; all calculators derive from the same base \texttt{Calculator} class, and comprise of four essential methods: \texttt{GetContract()}, \texttt{Open()}, \texttt{Process()} and \texttt{Close()}.

Calculator authors can specify the expected types of inputs and outputs of a calculator in \texttt{GetContract()}. When a graph is initialized, the framework calls a static method to verify if the packet types of the connected inputs and outputs match the information in this specification.

After a graph starts, the framework calls \texttt{Open()}. The input side packets are available to the calculator at this point. \texttt{Open()} interprets the node configuration (see Section \ref{graph_config}) operations and prepares the calculator's per-graph-run state. This function may also write packets to calculator outputs. An error during \texttt{Open()} can terminate the graph run.

For a calculator with inputs, the framework calls \texttt{Process()} repeatedly whenever at least one input stream has a packet available. The framework by default guarantees that all inputs have the same timestamp (see Section \ref{scheduling} for more information). Multiple \texttt{Process()} calls can be invoked simultaneously when parallel execution is enabled. If an error occurs during \texttt{Process()}, the framework calls \texttt{Close()} and the graph run terminates.

After all calls to \texttt{Process()} finish or when all input streams close, the framework calls \texttt{Close()}. This function is always called if \texttt{Open()} was called and succeeded and even if the graph run terminated because of an error. No inputs are available via any input streams during \texttt{Close()}, but it still has access to input side packets and therefore may write outputs.\footnote{A use case here is a media decoder reaching the end of file but still having additional images in its encoding state.} After \texttt{Close()} returns, the calculator should be considered a dead node. The calculator object is destroyed as soon as the graph finishes running.

\subsection{Graph}

In MediaPipe, all processing takes places within the context of a \texttt{Graph}. A graph contains a collection of nodes joined by directed connections along which packets can flow. Various resources required for the execution of nodes, such as the scheduler (further discussed in Section \ref{scheduling}), are also attached to the graph. A graph is typically defined via a graph configuration as a separate file or can be built programmatically in code.

In a graph, data flow can originate from \textit{source} nodes which have no input streams and produce packets spontaneously (\eg, by reading from a file). Data flow can also originate from graph input streams which allow an application to feed packets into a graph (\eg, passing in camera texture obtained from the operating system). Similarly, there are \textit{sink} nodes that receive data and write it to various destinations (\eg, a file, a memory buffer, etc.). An application can also receive outputs from the graph using callbacks or poll any output streams via output stream polling functions.

When a graph is initialized, the following constraints are checked:
\vspace{-\topsep}
\begin{enumerate}
\setlength{\parskip}{0pt}
\setlength{\itemsep}{0pt plus 1pt}
    \item Each stream and side packet must be produced by one source.
    \item The type of an input stream/side packet must be compatible with the type of the output stream/side packet to which it is connected.
    \item Each node's connections are compatible with its contract. 
\end{enumerate}
\vspace{-\topsep}
The function returns an error if the graph fails the validation step. 

During a single graph run, the framework constructs calculator objects corresponding to a graph node and calls \texttt{Open()}, \texttt{Process()} and \texttt{Close()} methods on these objects as discussed in {Section \ref{calculators}}.
The graph can stop running when:
\vspace{-\topsep}
\begin{itemize}
\setlength{\parskip}{0pt}
\setlength{\itemsep}{0pt plus 1pt}
    \item \texttt{Calculator::Close()} has been called on all calculators, or
    \item All source calculators indicate that they have finished sending packets and all graph input streams have been closed, or
    \item Any error occurs (the graph returns an error with a message in this case).
\end{itemize}

\subsection{GraphConfig} \label{graph_config}

A \texttt{GraphConfig} is a specification that describes the topology and functionality of a MediaPipe graph.

In the specification, a node in the graph represents an instance of a particular calculator. All the necessary configurations of the node, such its type, inputs and outputs must be described in the specification. Description of the node can also include several optional fields, such as node-specific options, input policy and executor, discussed in Section \ref{scheduling}.


To modularize a \texttt{GraphConfig} into sub-modules and assist with re-use of perception solutions, a MediaPipe graph can be defined as a \texttt{Subgraph}. The public interface to a subgraph consists of a set of input and output streams similar to the public interface of a calculator. The subgraph can then be included in an \texttt{GraphConfig} as if it were a calculator. When a MediaPipe graph is loaded from a \texttt{GraphConfig}, each subgraph node is replaced by the corresponding graph of calculators. As a result, the semantics and performance of the subgraph is identical to the corresponding graph of calculators.


\texttt{GraphConfig} has several other fields to configure the global graph-level settings, \eg, graph executor configs, number of threads, and maximum queue size of input streams. Several graph-level settings are useful for tuning the performance of the graph on different platforms (\eg, desktop v.s. mobile). For instance, on mobile, attaching a heavy model-inference calculator to a separate executor can improve the performance of a real-time application since this utilizes thread locality.

\section{Implementation} \label{implementation}


This section discusses MediaPipe's scheduling logic and powerful synchronization primitives to process time-series in a customizable fashion.

\subsection{Scheduling} \label{scheduling}

\subsubsection{Scheduling mechanics} \label{scheduling_mechanics}

Data processing in a MediaPipe graph occurs inside Calculators (nodes). The scheduling system decides when calculator code is ready and should be run.

Each graph has at least one scheduler queue. Each scheduler queue has exactly one executor. Nodes are statically assigned to a queue (and therefore to an executor). By default there is one queue whose executor is a thread pool with a thread count based on the system's capabilities.

Each node has a scheduling state which can be either \textit{not ready}, \textit{ready} or \textit{running}. A readiness function determines whether a node is ready to run, as discussed below. This function is invoked at graph initialization, also when a node finishes running or when the state of an input to a node changes.

The readiness function used depends on the type of node. Source nodes (\ie, nodes with no input streams) are always ready to run until they inform the framework that they have no more data to provide, at which point they are closed.

Non-source nodes are ready if they have inputs to process, and if those inputs form a valid input set according to the conditions described by the node's \textit{input policy} (\eg all inputs with matching timestamps are available, discussed in detail below). Most nodes use the default input policy, but developers can specify a different one. \footnote{Changing the input policy changes what guarantees the calculator's code can expect from its inputs. It is generally not possible to mix and match calculators with arbitrary input policies. Thus, a calculator that uses a special input policy should be written for it, and declare it in its contract.}

When a node becomes ready for execution, a task is added to the corresponding scheduler queue, which is a priority queue. When the graph is initialized, nodes are topologically sorted and assigned a priority based on the graph's layout; for example, nodes closer to the output side of the graph have higher priority, while source nodes have the lowest priority.

Executors are responsible for actually running the task by invoking the calculator's code. By configuring different executors the developer can gain finer control on the use of execution resources; for example, assigning certain nodes to run on lower priority threads.

\subsubsection{Synchronization} \label{synchronization}

MediaPipe graph execution is decentralized: there is no global clock, and different nodes can process data from different timestamps at the same time. This allows higher throughput via pipelining.

However, time information is very important for many perception workflows. Nodes that receive multiple input streams generally need to coordinate them in some way. For example, an object detector may output a list of boundary rectangles from a frame, and this information may be fed into a rendering node, which should process it together with the original frame.

Therefore, one of the key responsibilities of the MediaPipe framework is to provide input synchronization for nodes. In terms of framework mechanics, the \textit{primary} role of a timestamp is to serve as a \textit{synchronization key}.

Furthermore, MediaPipe is designed to support deterministic operations, which is important in many scenarios (testing, simulation, batch processing, etc.), while allowing graph authors to relax determinism where needed to meet real-time constraints.

The two objectives of synchronization and determinism underlie several design choices. Notably, the packets pushed into a given stream must have monotonically increasing timestamps\footnote{This constraint is applied at the level of the individual stream: there is no global timestamp.}: this is not just a useful assumption for many nodes, but it is also relied upon by our synchronization logic. Each stream has a \textit{timestamp bound}, which is the lowest possible timestamp allowed for a new packet on the stream. When a packet with timestamp $T$ arrives, the bound automatically advances to $T+1$, reflecting the monotonic requirement. This allows the framework to know for certain that no more packets with timestamp $< T$ will arrive.

\subsubsection{Input policies}

Synchronization is handled locally on each node, using the input policy specified by the node.

The \textit{default input policy} provides deterministic synchronization of inputs, with the following guarantees:
\vspace{-\topsep}
\begin{itemize}
\setlength{\parskip}{0pt}
\setlength{\itemsep}{0pt plus 1pt}
    \item If packets with the same timestamp are provided on multiple input streams, they will always be processed together regardless of their arrival order in real time.
    \item Input sets are processed in strictly ascending timestamp order. \footnote{An important consequence of this is that if the calculator always uses the current input timestamp when outputting packets, the output will inherently obey the monotonically increasing timestamp requirement.}
    \item No packets are dropped, and the processing is fully deterministic.
    \item The node becomes ready to process data as soon as possible given the guarantees above.
\end{itemize}
\vspace{-\topsep}

To explain how it works, we introduce the definition of a \textit{settled timestamp}. We say that a timestamp in a stream is \textit{settled} if it is lower than the timestamp bound. In other words, a timestamp is settled for a stream once the state of the input at that timestamp is irrevocably known: either there is a packet, or it is certain that a packet with that timestamp will not arrive. \footnote{For this reason, MediaPipe also allows a stream producer to explicitly advance the timestamp bound farther that what the last packet implies, i.e. to provide a tighter bound. This can allow the downstream nodes to settle their inputs sooner.}

Given this definition, a calculator with the default input policy is ready if there is a timestamp which is settled across all input streams and contains a packet on at least one input stream. The input policy provides all available packets for a settled timestamp as a single \textit{input set} to the calculator.\footnote{Note that it is \textit{not} guaranteed that an input packet will always be available for all streams.}

\begin{figure}[t]
\begin{center}
   \includegraphics[width=0.65\linewidth]{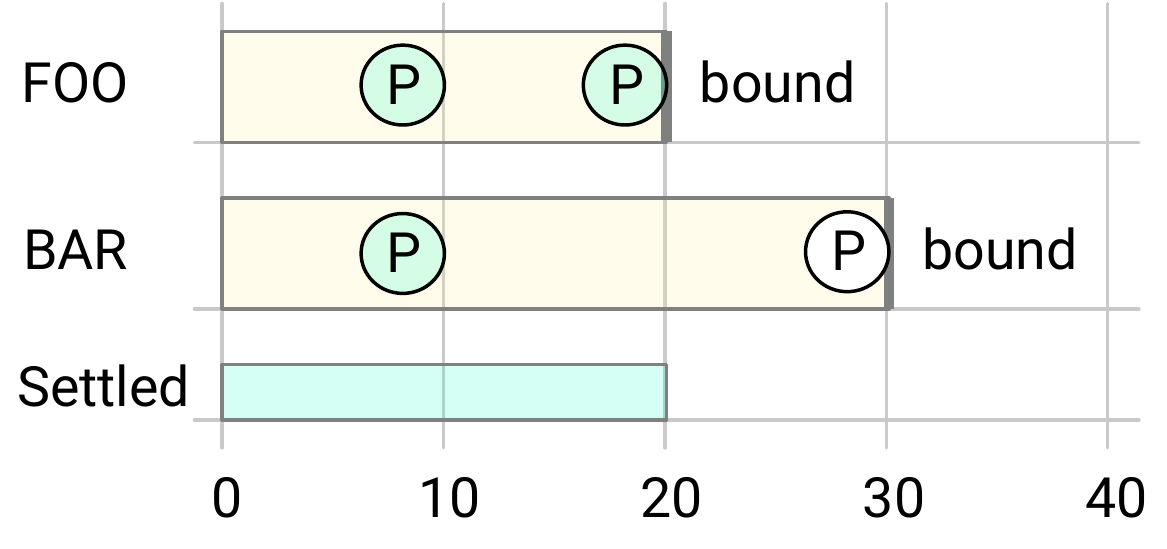}
\end{center}
   \caption{Input policy example. FOO and BAR are input streams, P denotes a packet, packets with settled timestamps shown in green. See text for details.}
\label{fig:input_policy}
\end{figure}

For example, see Figure \ref{fig:input_policy}. A node has two input streams, \texttt{FOO} and \texttt{BAR}. At this time, the node has received packets on \texttt{FOO} at timestamps $10$ and $20$, and on \texttt{BAR} at timestamps $10$ and $30$. All timestamps up to $20$ are settled, and the node's \texttt{Process()} method can be invoked for timestamp $10$ (with a packet for \texttt{FOO} and one for \texttt{BAR}) and for timestamp $20$ (with a packet for \texttt{FOO}, and no packet for \texttt{BAR}). However, timestamp $30$ cannot be processed yet, because the state of FOO is unknown past $20$. For instance, if \texttt{FOO} sends a packet with timestamp $25$, it will have to be processed before $30$ can be processed.

Besides the default synchronization policy, we allow nodes to specify other input policies. For example, a node can choose to receive all inputs immediately (sacrificing several of the guarantees listed earlier), or group its inputs into separate sets, enforcing timestamp synchronization only within but not across sets.

\subsubsection{Flow control}

Since packets may be generated faster than they can be processed, flow control is necessary to keep resource usage under control. Two mechanisms are available: a simple back-pressure system and a richer node-based system.

The back-pressure mechanism throttles the execution of upstream nodes when the packets buffered on a stream reach a limit. This mechanism maintains deterministic behavior and includes a deadlock avoidance system that relaxes configured limits when needed. It is suitable for reducing resource usage in batch operations.

The second system consists of inserting special nodes which can drop packets according to real-time constraints. Typically, these nodes use special input policies to be able to make fast decisions on their inputs. This node-based approach lets the graph author control where packets can be dropped, and allows flexibility in adapting and customizing the graph’s behavior depending on resource constraints.

\begin{figure}[t]
\begin{center}
   \includegraphics[width=0.95\linewidth]{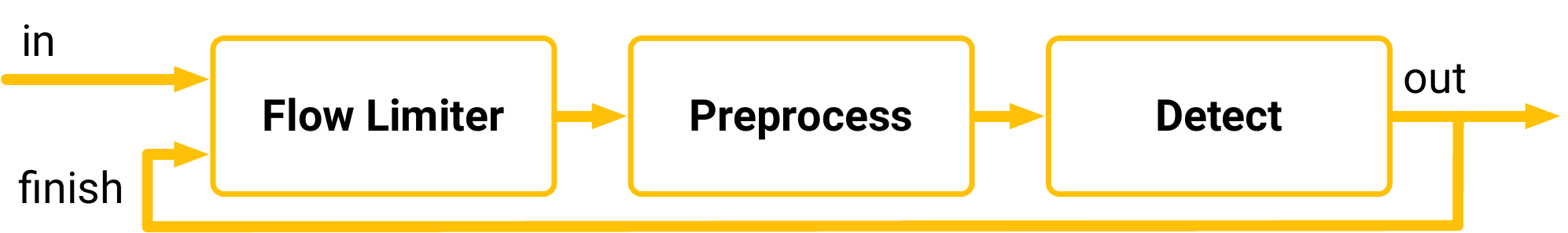}
\end{center}
   \caption{Flow limiter example. See text for details.}
\label{fig:flow_control}
\end{figure}

For example, Figure  \ref{fig:flow_control} shows a common pattern that places a flow-limiter node at the input of a subgraph, with a loopback connection from the final output to the flow-limiter. The flow-limiter is thus able to keep track of how many timestamps are being processed in the downstream graph, and can drop packets if this count hits a limit. Since packets are dropped upstream, we avoid the wasted work that would result from partially processing a timestamp and then dropping packets between intermediate stages. 

\subsection{GPU support}

MediaPipe supports GPU compute and rendering nodes, and allows combining multiple GPU nodes, as well as mixing them with CPU based nodes. There exist several GPU APIs on mobile platforms (\eg, OpenGL ES, Metal and Vulkan). MediaPipe does not attempt to offer a single cross-API GPU abstraction. Individual nodes can be written using different APIs, allowing them to take advantage of platform specific features when needed.
Our GPU support enables GPU nodes to enjoy the same advantages of encapsulation and composability as CPU nodes, while maintaining efficiency.

\subsubsection{Opaque buffer type}

GPU nodes in MediaPipe use an opaque buffer type to represent data accessible by the GPU (\eg, a video frame). This opaque type has multiple concrete implementations (\eg, depending on the platform). When a node wants to access the buffer using some API, it uses a helper class to obtain an API-specific view of the buffer, \eg, an OpenGL texture. This view object is ephemeral, and is released as soon as the node is done with its current processing task.

The creation and destruction of this view object provides the system with an opportunity to perform required tasks, to bind the data for the desired API and for synchronization.

\subsubsection{OpenGL support}

MediaPipe supports OpenGL ES up to version 3.2 on Android and up to ES 3.0 on iOS. In addition, MediaPipe also supports Metal on iOS. We discuss the OpenGL implementation on Android in more detail here.

MediaPipe allows graphs to run OpenGL in multiple GL contexts. For example, this can be very useful in graphs that combine a slower GPU inference path (\eg, at 10 FPS) with a faster GPU rendering path (\eg, at 30 FPS): since one GL context corresponds to one sequential command queue, using the same context for both tasks would reduce the rendering frame rate. One challenge MediaPipe's use of multiple contexts solves is the ability to communicate across them. An example scenario is one with an input video that is sent to both the rendering and inferences paths, and rendering needs to have access to the latest output from inference.

An OpenGL context cannot be accessed by multiple threads at the same time. Furthermore, switching the active GL context on the same thread can be slow on some Android devices. Therefore, our approach is to have one dedicated thread per context. Each thread issues GL commands, building up a serial command queue on its context, which is then executed by the GPU asynchronously.

Multiple OpenGL contexts can share resources (\eg, textures) if they are properly connected when they are created. However, OpenGL does not synchronize the state of these objects across contexts. For example, if a texture is written to in context A and then read from in context B, the read operation may not see the updates made by context A. It is therefore not sufficient to introduce synchronization between the CPU threads serving the two contexts - even if the CPU issues the write commands for context A before the read commands for context B, there is no guarantee that the GPU will obey the same ordering when executing the two command sets.

OpenGL offers \textit{sync fence} objects as a low-level mechanism for cross-context synchronization. A sync fence can be created in context A's command stream, and context B can then insert a \textit{wait} operation on A's fence in its own command stream. This ensures commands in B's command queue following the wait will only be executed once all commands in A's queue predating the sync fence finished execution on the GPU.

Managing this normally requires additional effort from the programmer. MediaPipe reduces the workload by automatically inserting synchronization operations in the GPU command streams where appropriate.

For each buffer, we keep track of one \textit{producer} sync fence, and multiple \textit{consumer} sync fences. The producer fence is inserted when the producer node outputs a buffer, after the commands used to write the buffer's contents; in other words, it marks a ``write complete'' point. When consumers on a different GL context request read access to the buffer, a wait operation on the producer fence is inserted.
Consumer fences are inserted when a consumer node is done with a buffer, and mark ``read complete'' points. These are used when the buffer is recycled:  before passing it to a new producer for writing, the framework waits for all existing consumers to finish reading the old contents.

Note that synchronization is done in the GPU command stream whenever possible, without forcing a CPU sync. The framework tries to avoid GPU/CPU sync operations as much as possible, and allows pipelining between GPU and CPU tasks.

\section{Tools} \label{tools}

This section describes some developer tools that help MediaPipe users analyze the performance of their perception pipelines.

\subsection{Tracer} \label{tracer}

The MediaPipe tracer module follows individual packets across a graph and records timing events along the way. With each event, it records a \texttt{TraceEvent} structure with several data fields \texttt{event\_time}, \texttt{packet\_timestamp}, \texttt{packet\_data\_id}, \texttt{node\_id}, and \texttt{stream\_id}. This \texttt{TraceEvent} information is sufficient to follow the flow of data and execution across the graph. The tracer also reports histograms of various resources, such as the elapsed CPU time across each calculator and across each stream.

The timing data recorded by the tracer module enables reporting and visualization of individual packet flows and individual calculator executions. The recorded timing data can be used to diagnose a variety of problems, such as unexpected real-time delays, memory accumulation due to packet buffering, and collating packets at different frame rates. The timing data can also be aggregated to report average and extreme latencies which is useful for performance tuning. Furthermore, the timing data can be explored to identify the calculators along the \textit{critical path}, whose performance determines end-to-end latency.


The tracer module records timing information on demand, and can be enabled using a section of the \texttt{GraphConfig}. Also, a MediaPipe user can completely omit the tracer module code using a compiler flag. The tracer module internally stores \texttt{TraceEvent} records in a circular buffer. In order to avoid thread contention over the circular buffer and to minimize the impact on timing measurements, the tracer module utilizes a mutex-free thread-safe buffer implementation.


\subsection{Visualizer}

\begin{figure}[t]
\begin{center}
   \includegraphics[width=\linewidth, frame]{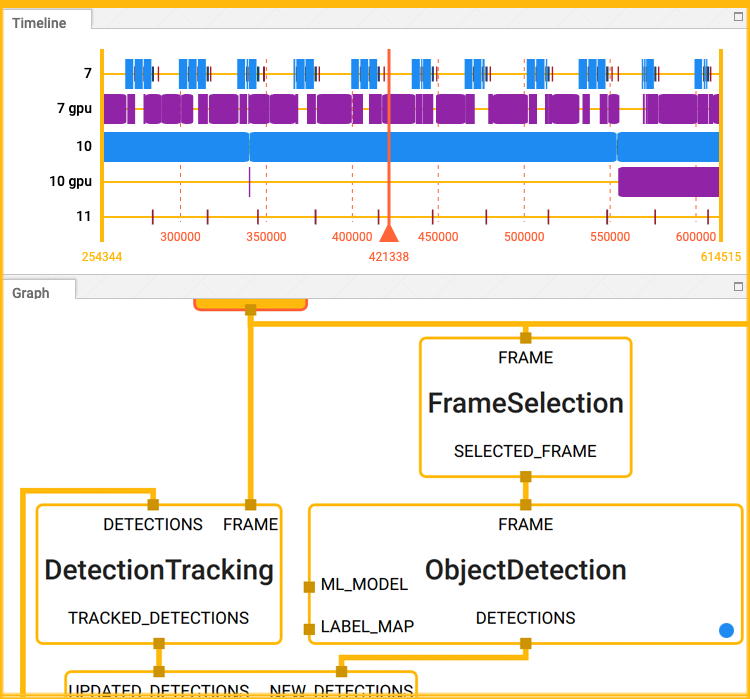}
\end{center}
   \caption{Visualizer example: The top half shows the Timeline view illustrating the timing of packets for each thread (row). The bottom half shows the Graph view illustrating how the different calculators are connected together.}
\label{fig:viz}
\end{figure}

MediaPipe visualizer is a tool that helps users understand the topology and overall behavior of their pipelines, as shown in Figure \ref{fig:viz}. The tool consists visualization of the following: 

\vspace{-\topsep}
\begin{itemize}
\setlength{\parskip}{0pt}
\setlength{\itemsep}{0pt plus 1pt}
\item Timeline view: A user can load a pre-recorded trace file (see Section \ref{tracer}) and see the precise timing of packets as they move through threads and calculators.
\item Graph view: A user can also visualize the topology of a graph as inferred from the same trace file that drives the Timeline view. This lets the user observe the full state of the graph at any point in time, including the state of each calculator and the packets being processed or being held in its input queues.
\end{itemize}

\section{Application examples} \label{application}

In this section, we discuss two different pipelines built with MediaPipe.

\subsection{Object detection} \label{object}

Real-time object detection from a live camera feed is a common perception application. Depending on the target device platform, running ML-based object detection at a full frame rate (\eg, 30 FPS) can require high resource consumption or be potentially infeasible due to long inference times. An alternative is to apply object detection to a temporally sub-sampled stream of frames and propagate the detection results, \ie, bounding boxes and the corresponding class labels, to all frames using a lightweight tracker. For optimal performance, tracking and detection should be run in parallel, so the tracker is not blocked by the detector and can process every frame.

This perception pipeline can be easily implemented with MediaPipe, as presented in the example graph in Figure \ref{fig:objectgraph}. There are two branches in the beginning of the graph: one slow branch for detection and one fast branch for tracking. Calculators for these tasks can be configured to run on parallel threads with the specification of executors in the pipeline's graph configuration (refer to Section \ref{graph_config} and \ref{scheduling_mechanics}). In the detection branch, a frame-selection node first selects frames to go through detection based on limiting frequency or scene-change analysis, and passes them to the detector while dropping the irrelevant frames. The object-detection node consumes an ML model and the associated label map as input side packets, performs ML inference on the incoming selected frames using an inference engine (\eg, \cite{tflite} or \cite{caffe2}) and outputs detection results.

In parallel to the detection branch, the tracking branch updates earlier detections and advances their locations to the current camera frame.

After detection, the detection-merging node compares results and merges them with detections from earlier frames removing duplicate results based on their location in the frame and/or class proximity.

Note, that the detection-merging node operates on the same frame that the new detections were derived from. This is automatically handled by the default input policy in this node as it aligns the timestamps of the two sets of detection results before they are processed together (see Section \ref{synchronization} for more information). The node also sends merged detections back to the tracker to initialize new tracking targets if needed.

For visual display, the detection-annotation node adds overlays with the annotations representing the merged detections on top of the camera frames, and the synchronization between the annotations and camera frames is automatically handled by the default input policy before drawing takes place in this calculator. The end result is a slightly delayed viewfinder output (\eg, by a few frames) that is perfectly aligned with the computed and tracked detections, effectively hiding model latency in a dynamic way.

Since MediaPipe provides cross-platform support, as an example development flow, this graph can be first developed and tested on desktop followed by deployment and final performance evaluation on mobile devices. Moreover, with minimal changes to the graph specification and the associated data flow, a node in the graph can be replaced by another node with a similar purpose but a different implementation. For instance, a heavy NN-based object detector may be swapped out with a light template matching detector, and the rest of the graph can stay unchanged.

\subsection{Face landmark detection and segmentation} \label{face}

\begin{figure}[t]
\begin{center}
    \includegraphics[width=1.0\columnwidth]{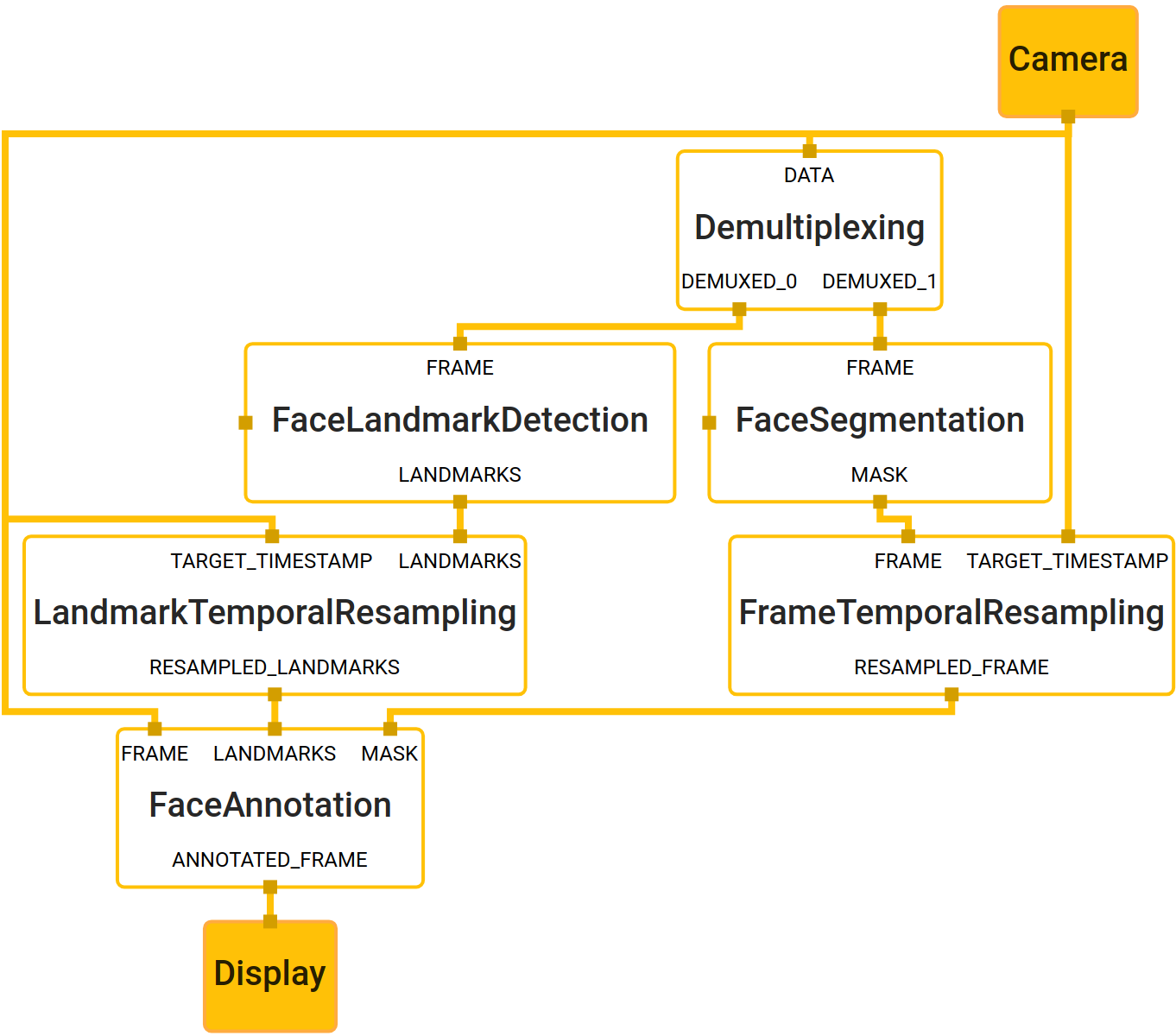}
    \caption{Face landmark detection and segmentation using MediaGraph.}
    \label{fig:facegraph_a}
\end{center}
\end{figure}

\begin{wrapfigure}{r}{0.35\columnwidth}
\begin{center}
    \includegraphics[width=0.3\columnwidth]{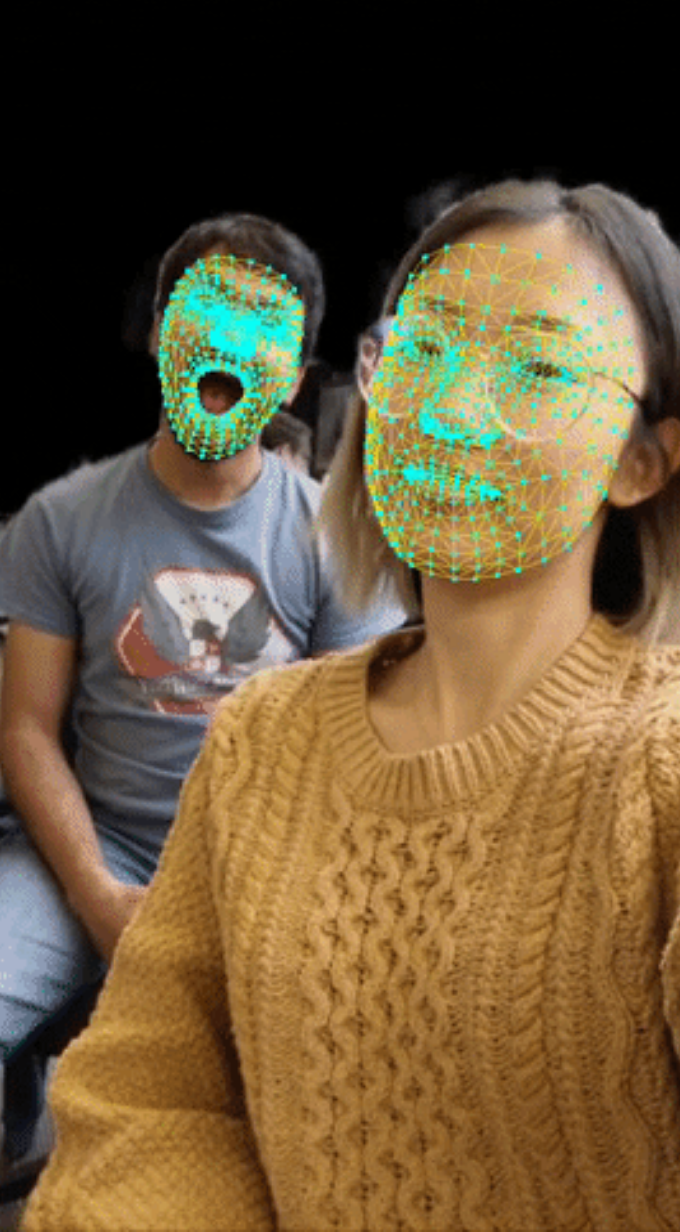}
        \caption{Landmark detection and segmentation output (\texttt{ANNOTATED\_FRAME})}
    \label{fig:facegraph_b}
\end{center}
\end{wrapfigure}

Face landmark estimation is another common perception application. Figure \ref{fig:facegraph_a} depicts a MediaPipe graph that performs face landmark detection along with portrait segmentation. To reduce the computational load needed to run both tasks simultaneously, one strategy is to apply the tasks on two disjoint subsets of frames. This can be done easily in MediaPipe using a demultiplexing node that splits the packets in the input stream into interleaving subsets of packets, with each subset going into a separate output stream.

To derive the detected landmarks and segmentation masks on all frames, the landmarks and masks are temporally interpolated across frames. The target timestamps for interpolation are simply those of all incoming frames. Finally, for visualization the annotations from the two tasks are overlaid onto the camera frames, with synchronization across the three streams handled by the input policy of the annotation node. A snapshot of the visual annotation is shown in Figure \ref{fig:facegraph_b}.

The pipeline can be further accelerated with GPU compute while reusing most of the pipeline configuration. For example, the face-landmark-detection node can switch to a GPU-based implementation, using a GPU inference engine (\eg, \cite{tflitegpu}). Additionally, temporal re-sampling and annotation can also have a GPU-based implementation. Together with the GPU support embedded in the framework, the entire data flow and compute can stay in GPU end-to-end with no speed bottlenecks commonly observed from GPU-to-CPU data transfer. Furthermore, it is also straightforward to configure a pipeline where the detection branch is performed on GPU while in parallel the segmentation branch is running on CPU for potentially better overall performance.

\section{Conclusion}

In this paper, we presented MediaPipe, a framework for building a perception pipeline as a graph of reusable calculators.
We described how the framework manages calculator execution using a 
comprehensive scheduling system, enables support for GPU on multiple platforms, and provides tools for evaluating graph performance.
MediaPipe can help a developer prototype very quickly and run a perception application efficiently across multiple platforms.



MediaPipe has been immensely successful at Google
for over 6 years.  One of the main reasons for its success can be attributed to
the ecosystem of re-usable calculators and graphs.
Given this experience, our primary focus after the open-source release will be our community support including third-party development of calculators and curating a set of recommended calculators and graphs.
Furthermore, we will further improve tooling to make performance and quality evaluation easy for the users.



{\small
\bibliographystyle{ieee_fullname}
\bibliography{egbib}
}

\end{document}